# Design and Architecture of Airmod: An Active Noise Cancellation and Deep Learning System for Real-Time Respiratory Rate Estimation from Tracheal Sounds

Shang-Ran Huang[1], Fu-Shun Hsu[1,2,3], Ji-Chang Huang[1], Chuan Chen[1], Shou-Zen Fan[4,5], Chang-Fu Su[4], Eng-Che Huang[6], Yuan-Ren Cheng[1,*]

[1] Heroic Faith Medical Science, Co., Ltd., New Taipei, Taiwan
[2] Department of Internal Medicine, Chung Shan Medical University, Taichung, Taiwan
[3] Hospital Development Affairs Office, Chung Shan Medical University Hospital, Taichung, Taiwan
[4] Department of Anesthesiology, National Taiwan University Hospital, Taipei, Taiwan
[5] Department of Anesthesiology, En Chu Kong Hospital, New Taipei, Taiwan
[6] Department of Anesthesiology, Taipei City Hospital Yangming Branch, Taipei, Taiwan

[*] Corresponding author
infie.cheng@heroic-faith.com

## 1. Introduction

More and more clinical anesthesia is given to subjects undergoing a diagnostic or therapeutic procedure in the form of monitored anesthesia care (MAC) (1, 2) or procedural sedation (3, 4) in a non-operating room setting. A statistical report revealed that the percentage of non-operating room anesthesia (NORA) cases among all reported anesthesia cases increased from 28.3% to 35.9% between 2010 and 2014 in the US (5). However, MAC/procedural sedation involves underlying risks. Studies show that 29.0%–35.4% of subjects under procedural sedation experience respiratory adverse events (6), and about 0.5% of subjects under procedural sedation experience sentinel hypoxia events (7). An American Society of Anesthesiologists Closed Claims analysis revealed that the most common NORA claims were related to severe respiratory events leading to death or permanent brain damage (8). Sixty-eight percent of these respiratory event-related claims were judged preventable if better respiratory monitoring had been applied (8).

Pre-cordial or pre-tracheal auscultation with a stethoscope is a conventional technique recommended by many MAC/procedural sedation practice guidelines to monitor the subjects' ventilation (9-13). Some U.S. state regulations recommend or require tracheal auscultation with a stethoscope for respiratory monitoring during sedation in outpatient or dental clinics (14-17). However, continuous auscultation with a handheld stethoscope is difficult. In addition, ambient noise may interfere with auscultation interpretation (18), and unexpected loud noise may harm the practitioner's hearing. Moreover, even for a clinical professional, interpreting auscultated respiratory sounds only by hearing is difficult (19).

Computerized respiratory sound analysis based on signal processing, machine learning, and deep learning has revolutionized conventional auscultation (20-22). Algorithms have been proposed to automatically detect breath phases (or inhalations and exhalations) in the breath sounds. Algorithms have been proposed to automatically detect breath phases or inhalations and exhalations in the tracheal sounds (23, 24). However, these algorithms mainly focused on detecting the event in a fixed-length recording (usually 15 to 90 seconds); they did not address real-time respiratory rate estimation based on the streaming of breath sounds.

In this paper, we present the development of an Artificial Intelligence Respiratory MOnitoring Device (Airmod™, Heroic Faith, New Taipei, Taiwan), hereinafter referred to as the proposed device, which is FDA-cleared under 510(k) K242798, that overcomes the abovementioned limitations. We first established a continuous tracheal sound database, HF_Tracheal_V2 (https://gitlab.com/techsupportHF/hf_tracheal_v2). A multiclass convolutional bidirectional gated recurrent unit (CNN-BiGRU) model was trained on the basis of HF_Tracheal_V2 to detect inhalations and exhalations in the tracheal sounds of 15-sec inference windows. We developed essential post-processing to estimate the respiratory rate in real time from sequential, overlapping 15-sec inference windows. Ambient noise interference was mitigated using active noise control (ANC) (25, 26). Real-time tracheal sound broadcasting and spectrogram display were designed to help clinicians identify respiratory abnormalities (27).

## 2. Software Architecture and Essential Function Modules of the Proposed Device

### 2.1 Software Architecture

The proposed device is a software as a medical device. Fig 1 shows the software architecture, essential modules, and data flow. The proposed device must run on an Android device that has sufficient computing power to support real-time inference of the CNN-BiGRU model. The Android version is locked at 11.0 during development. GPU acceleration is not used in the proposed device.

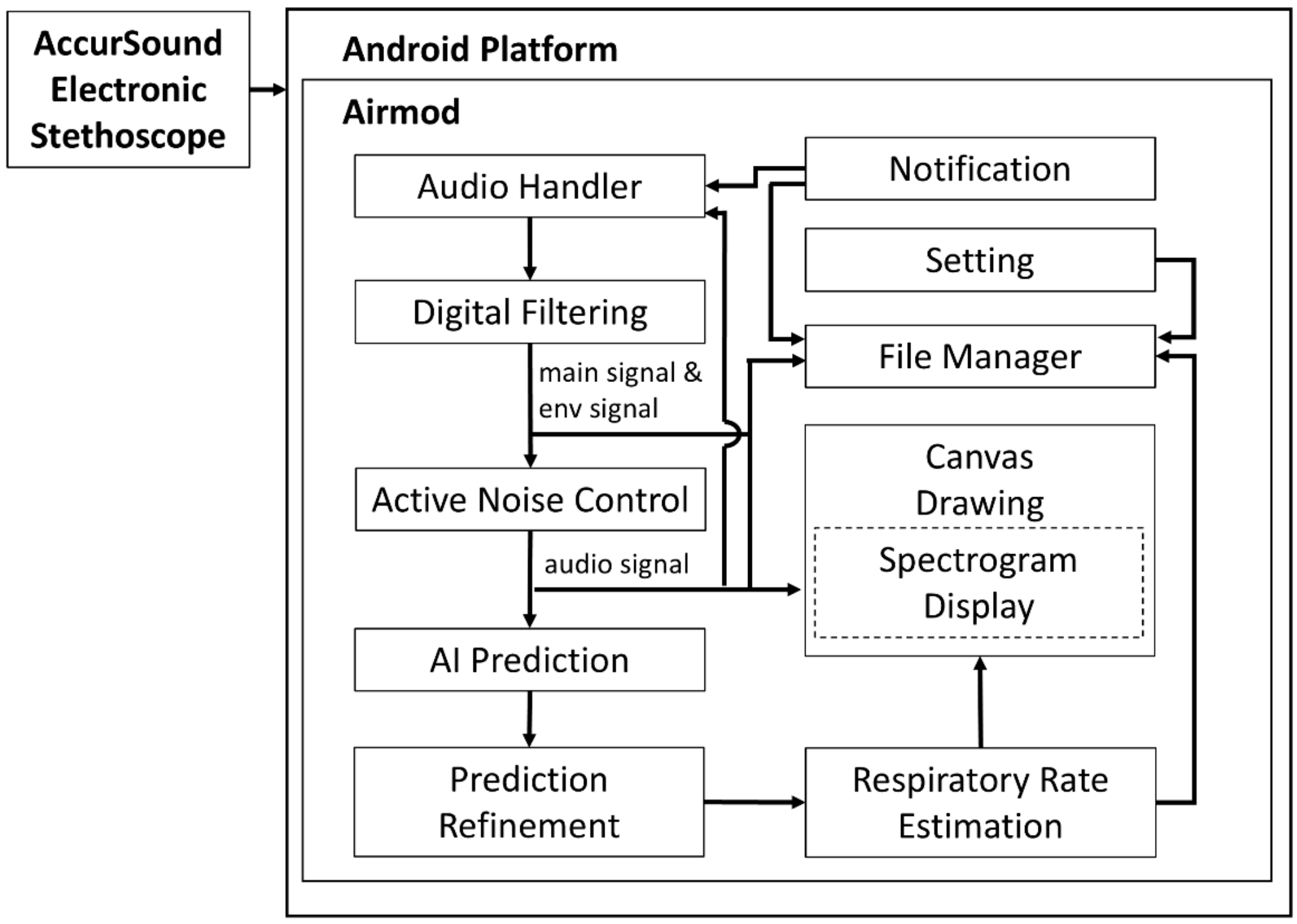

**Figure 1. Software architecture, essential modules, and data flow of the proposed device.** Black arrows indicate the data flow.

## 2.2 Essential Function Modules

### 2.2.1 Audio Handler

The module "Audio Handler" handles two jobs. Firstly, when the specified electronic stethoscope, AccurSound$^{TM}$ AS-101 (Heroic Faith, New Taipei, Taiwan), is connected to the proposed device, "Audio Handler" receives two sound signals, main and environmental (env), picked up by the acoustic sensors. The received signals are passed to the "Active Noise Control" module (explained in section 3). Secondly, the "Audio Handler" plays the processed sound signals and alarms to the Android device's built-in speaker or any connected audio output device.

### 2.2.2 Digital Filtering

In the module "Digital Filtering," a Butterworth band-pass filter with cut-off frequencies at 20 and 1500 Hz is applied to the signal to remove unwanted high-frequency noise.

### 2.2.3 Active Noise Control (ANC)

The "Active Noise Control" module uses a normalized least mean square (nLMS) adaptive finite impulse response (FIR) filter to cancel out the ambient noise entering the main mic (28). The ANC process is further discussed in section 3.

### 2.2.4 AI Prediction

In "AI Prediction," the proposed device prepares three sets of features: spectrogram, mel frequency cepstral coefficients (MFCCs) [20], and energy summations. These features are inputs to the CNN-BiGRU model, which predicts whether the signal's inhalation and exhalation are present in time segments. A postprocessing step merges the segments into events and removes unwanted segments. The details about the AI model are described in section 4.

### 2.2.5 Prediction Refinement

The inhalation events predicted by the CNN-BiGRU model may not be perfect. Furthermore, the model only deals with signals with a 15-second inference window (explained in section 5). Hence, the AI prediction results must be refined and concatenated before respiratory rate estimation. The process is clearly described and illustrated in section 5.

### 2.2.6 Respiratory Rate Estimation

The breath-to-breath respiratory rate is estimated from a history list that contains the start and end times of all refined and concatenated inhalation events. Because the respiratory rate is not updated until a new inhalation event is registered, we devised a strategy to make the respiratory rate change more smoothly when inhalation is not detected for some time. The strategy is clearly explained and illustrated in section 6.

### 2.2.7 File Manager

"File Manager" is responsible for archiving all the sound signals, keyed-in patient information, system information, and metadata. The main, env, and audio signals are continuously written into separate files in a WAVE format (.wav). The inhalation detection results are also written into a Text (.txt) file. The other information, such as metadata, is stored in a JSON (.json) file.

### 2.2.8 Canvas Drawing

"Canvas Drawing" is responsible for drawing the user interface and displaying the spectrogram and the respiratory rate in real time.

### 2.2.9 Notification

"Notification" monitors the estimated respiratory rate. If it drops below 4 breaths per minute (bpm), which means no inhalation event was detected in the last 15 seconds, the proposed device makes sounds to notify the clinicians.

### 2.2.10 Setting

"Setting" is a controller that handles the proposed device's user interaction. Through "Setting," users can configure the parameters in the other modules, key in patient information, turn on/off optional functions, adjust sound volume, etc.

## 3. Active Noise Control

Fig 2 illustrates the clinical scenario of respiratory sound monitoring. Ambient noise poses the greatest challenge to accurately estimating respiratory rate. The ambient noise may include, but is not limited to, music, sounds generated by medical instruments, devices, and equipment, sounds produced by venting systems and air-conditioning, and human voice. Ideally, a well-designed acoustic sensor should capture only respiratory sound (the green arrows in Fig 2), and its housing should prevent ambient noise from entering the sensor via the air path. However, the ambient noise can enter the sensor through the body path (the blue arrow in Fig 2) when it first hits the subject's body and travels through the body tissue. Ambient noise impairs auscultation, making it difficult to analyze auscultatory sounds; consequently, respiratory rate estimation may fail.

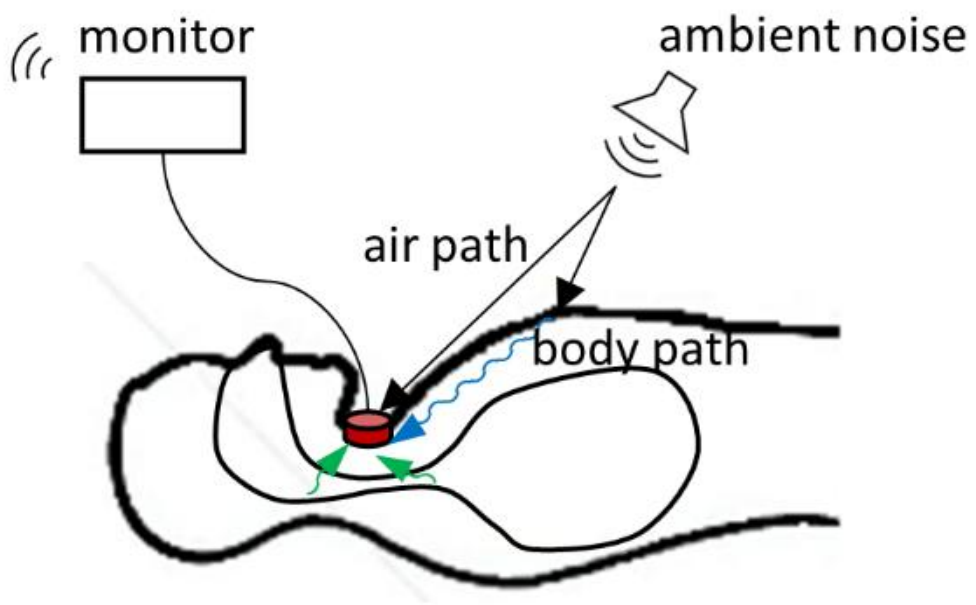

**Figure 2. Illustration of the clinical setting of the proposed device.** The red object represents the acoustic sensor (a microphone or a stethoscope chestpiece). The black arrows represent the ambient noise transmitted through the air. The blue arrow represents ambient noise that travels to the acoustic sensor through the body. The green arrows represent the tracheal sound transmission paths.

The proposed device uses a proprietary electronic stethoscope, AccurSound$^{TM}$ AS-101, to minimize ambient noise interference and employ ANC technology (28). The stethoscope sensor is equipped with dual microphones (mics), namely, main and env mics, aiming to pick up sound from two isolated air cavities. The dual-microphone design enables ANC. A normalized least-mean-square (nLMS) adaptive finite-impulse-response (FIR) filter is used in the proposed device to cancel ambient noise (28).

4. A Multiclass Deep Learning Model for Inhalation and Exhalation Detection

An algorithm that can identify inhalations and exhalations in tracheal sounds is required to enable the proposed device to estimate the respiratory rate. We previously trained CNN-BiGRU models to detect inhalation and exhalation, respectively, using the HF_Tracheal_V1 database (23). The training and test data in the HF_Trachal_V1 are truncated into 15-s clips. However, if a truncated 15-s clip contains no events, it is discarded to balance the foreground-to-background ratio. Discarding these clips makes reconstructing the original continuous recordings impossible, preventing us from developing strategies to manage continuous tracheal sound streaming. Hence, we established a new continuous tracheal sound database, HF_Tracheal_V2, and trained a new CNN-BiGRU model for the proposed device.

4.1 Study Participants

A total of 369 subjects from three hospitals who underwent a diagnostic/therapeutic procedure under MAC/procedural sedation were enrolled. The tracheal sound recording protocol was approved by the institutional review boards of Taiwan University Hospital, Taipei, Taiwan (IRB number: 202005108DIPC), Taipei City Hospital (IRB number: TCHIRB-10811009-E), and En Chu Kong Hospital (IRB number: ECKIRB1090303). It has also been further approved by the Joint Institutional Review Board of the Medical Research Ethical Foundation, Taipei, Taiwan (case number: 19-006-A-2). The study was conducted in accordance with the Helsinki Declaration and its following amendments.

Table 1 lists the demographic characteristics of the subjects. Their age, height, weight, and BMI were presented as mean ± standard deviation. Some data are missing because not all protocols require written informed consent.

**Table 1. Demographic characteristics of the subjects.**

| | |
|---|---|
| Number | 369 |
| Sex (M/F/unknown) | 123/188/58 |
| Age | 53.6 ± 14.0* |
| Height (cm) | 162.3 ± 7.9* |
| Weight (kg) | 65.0 ± 12.4† |
| BMI (kg/m$^2$) | 24.6 ± 3.9† |
| Procedure, n (%) | |
| Colostomy | 2 (0.5%) |
| Colon surgery | 2 (0.5%) |
| Laparoscopy | 1 (0.3%) |
| Endoscopic Ultrasound | |
| Stomach | 2 (0.5%) |
| Caesarean Section | |
| Extraperitoneal | 38 (10.3%) |
| Intraperitoneal | 10 (2.7%) |
| Unspecified | 5 (1.4%) |
| GI Endoscopy | |
| UGI | 5 (1.4%) |
| LGI | 71 (19.2%) |
| UGI + LGI | 90 (24.4%) |
| Unspecified | 141 (38.2%) |
| Unknown | 2 (0.5%) |

*56 missing data; †59 missing data. BMI: body mass index. GI: gastrointestinal. LGI: lower gastrointestinal. UGI: upper gastrointestinal.

4.2 Tracheal Sound Recording

The HF-Type-2 (Fig 3a) device, composed of the AccurSound AS-101 device and a smartphone (Mi 9T pro, Xiaomi, Beijing, China) with a custom app, was used to record the tracheal sounds of the subjects during the procedures. The acoustic sensor was attached to the flat area next to the subjects' thyroid cartilage (Fig 3b) and affixed with 3M tape. Continuous recording began before the induction dose of anesthetics and continued until the procedure was completed. The sampling rate was 4000 Hz. The AccurSound AS-101's analog-to-digital converter had a 16-bit bit depth; however, the app converted it to 32-bit before recording. The anesthetics/sedatives are given by bolus injection. The anesthetic/sedative drug or drug combinations administered to the subjects included: 1) remifentanil and propofol for intravenous anesthesia, 2) midazolam, alfentanil, and propofol for intravenous anesthesia, 3) bupivacaine for spinal anesthesia, or 4) lidocaine/levobupivacaine/ropivacaine for epidural anesthesia. A total of 373 tracheal sound recordings were collected from the subjects. The sounds were collected between March 2020 and January 2021.

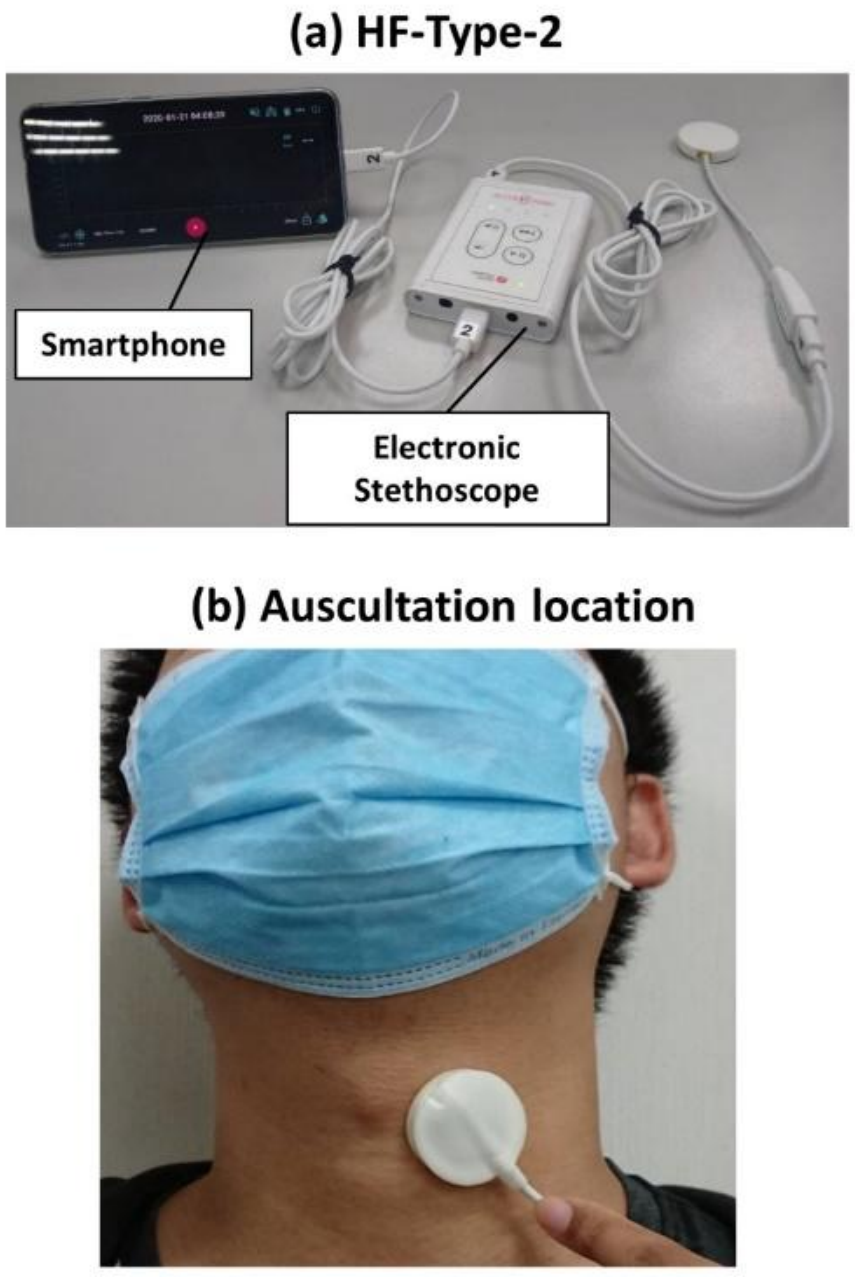

Figure 3. (a) the HF-Type-2 device, (b) the auscultation location at the flat area next to the thyroid cartilage. These photos are reproduced from our previous study (23) under a Creative Commons (CC BY) international license.

During the recording, supplemental oxygen was administered to the subjects via nasal cannula. Most of the time, the oxygen flow rate provided to the subjects was ≤3 L/min; however, the anesthesiologists may adjust the flow rate based on the subjects' condition. Basic airway maneuvers, such as chin lift and jaw thrust, may be performed on subjects if they develop partial airway obstruction or apnea.

4.3 Data Labeling

Five clinical experts, including three respiratory therapists and two nurses, were recruited to label the tracheal sounds. The Examination Yuan of Taiwan issues their professional licenses. They practiced their profession in Taiwan. They had an average of 12.8 ± 3.27 years of clinical experience. They labeled the recordings using custom labeling software (29). They were taught to first listen for the tracheal sound, then, if necessary, observe the corresponding spectrogram several times before labeling the start and end times of inhalation, exhalation, and continuous adventitious sounds (CAS). They were not asked to distinguish among snoring, stridor, rhonchi, and wheezes; all of these sounds were categorized as CAS. They can fast-forward or go back to any point in a tracheal sound recording to correct a previous label at any time. Two other respiration therapists, with 6 and 3 years of clinical experience, reviewed the labels and may return the files to the labelers to correct incorrect labels. A regular meeting was held to ensure the labelers agreed on the labeling criteria. Note that the inhalation (I), exhalation (E), and CAS labels (C) were still not a perfect ground truth. This study did not use the C labels.

### 4.4 Data Set Assignment

We randomly and subject-wise distributed the recordings to either the training or test set of HF_Tracheal_V2. We set the ratio of tracheal sound recordings in the training set to the test set to approximately 4:1.

### 4.5 Preparing Training Data

Because the original duration of the tracheal sound recordings in HF_Tracheal_V2 ranged from a few minutes to a few hours, it wasn't easy to have a model analyzing such long signals in one sitting. Thus, we used a 15-s sliding window to truncate the continuous recordings to prepare the training data. The truncation length was decided to be 15 s for the following reasons: 1) 3−5 breaths are usually contained in 15 s, which is long enough for a clinician to make a clinical decision; 2) a previous study used 15-s sound to standardize the nomenclature in respiratory auscultation (30), and 3) an apnea event is defined as ≥10-s or ≥15-s cessation of breathing (31, 32). The stride of the sliding window was set to 5-s for data augmentation during training. The file was not used for training if a 15-s clip did not contain a label.

### 4.6 Feature Preparation

Fig 4 displays the deep learning pipeline used in this study. In the feature preparation stage, we followed the same approach in our previous study (23) to prepare the features fed into the input layers of the CNN-BiGRU models. A Butterworth high-pass filter was first applied to the raw 15-s tracheal sound, with a cutoff frequency of 80 Hz. Then, the 938×129 spectrogram, 938×60 MFCCs (33), and 938×4 energy summation vectors were calculated. The spectrogram was derived by applying a short-time Fourier transform (window type: Hanning, window width: 256, hop length: 64, and zero-padding: none) to the 15-s truncated tracheal sound. The MFCCs, including 938×20 static, 938×20 delta, and 938×20 delta-delta matrices, were computed from the spectrogram. The four energy summation vectors were computed by summing the energy within the four frequency bands, namely, 0–250, 250–500, 500–1,000, and 0–2,000 Hz, along the frequency axis of the spectrogram. Finally, min-max normalization was applied to the spectrogram, each of the three MFCC matrices, and each of the energy summation vectors.

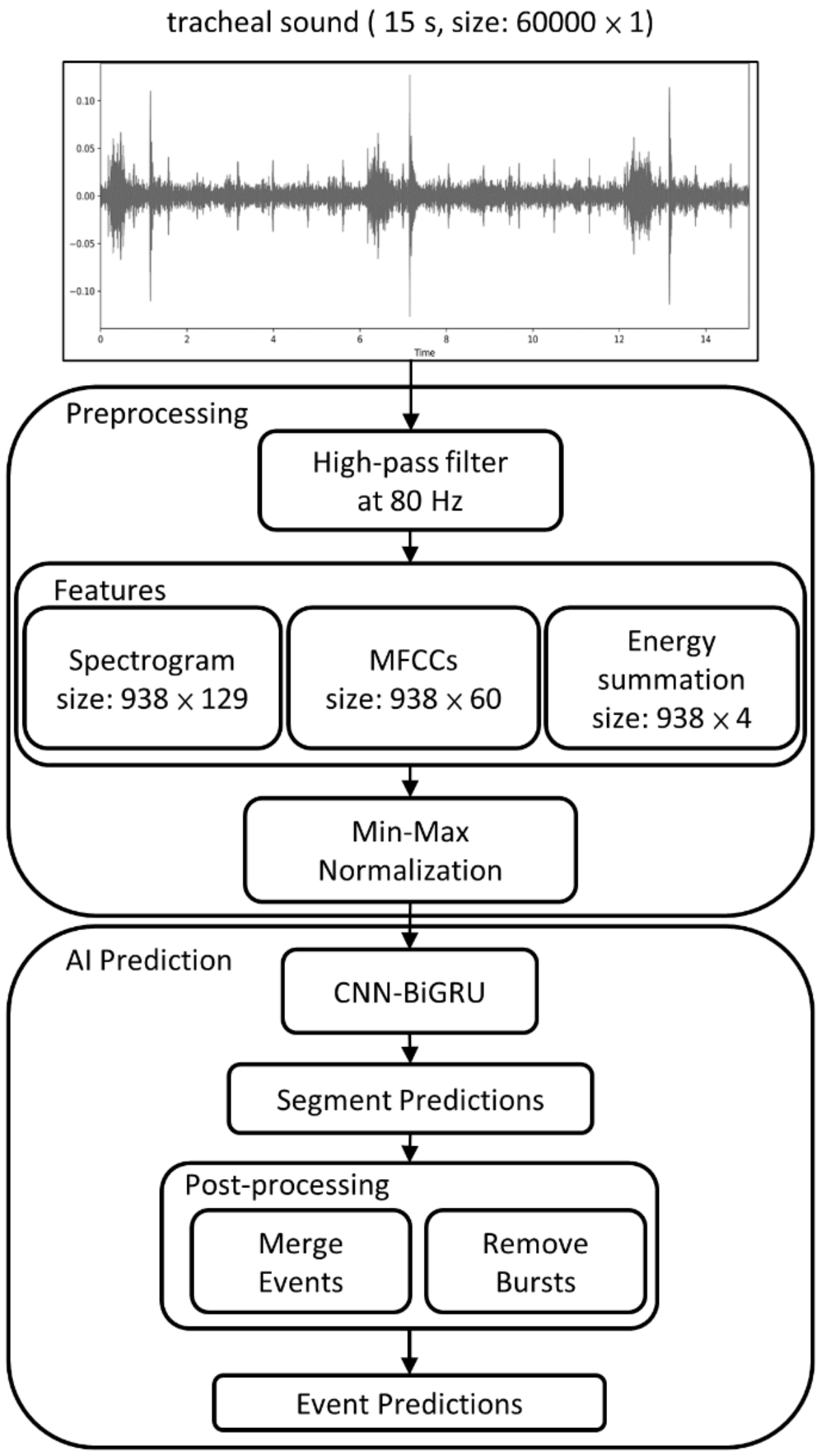

**Fig 4. The deep learning pipeline**. This graph is reproduced from our previous study (23) under a Creative Commons 4.0 (CC BY 4.0) international license.

4.7 The Architecture of the Deep Learning Model

Fig 5 displays the architecture of the CNN-BiGRU model. In this study, we made minor modifications to the model previously used in lung and tracheal sound analysis (23). We fed the three features (spectrogram, MFCCs, and energy summation) into three separate paths. Moreover, we used the hyperbolic tangent (tanh) function as the output layer rather than the sigmoid function.

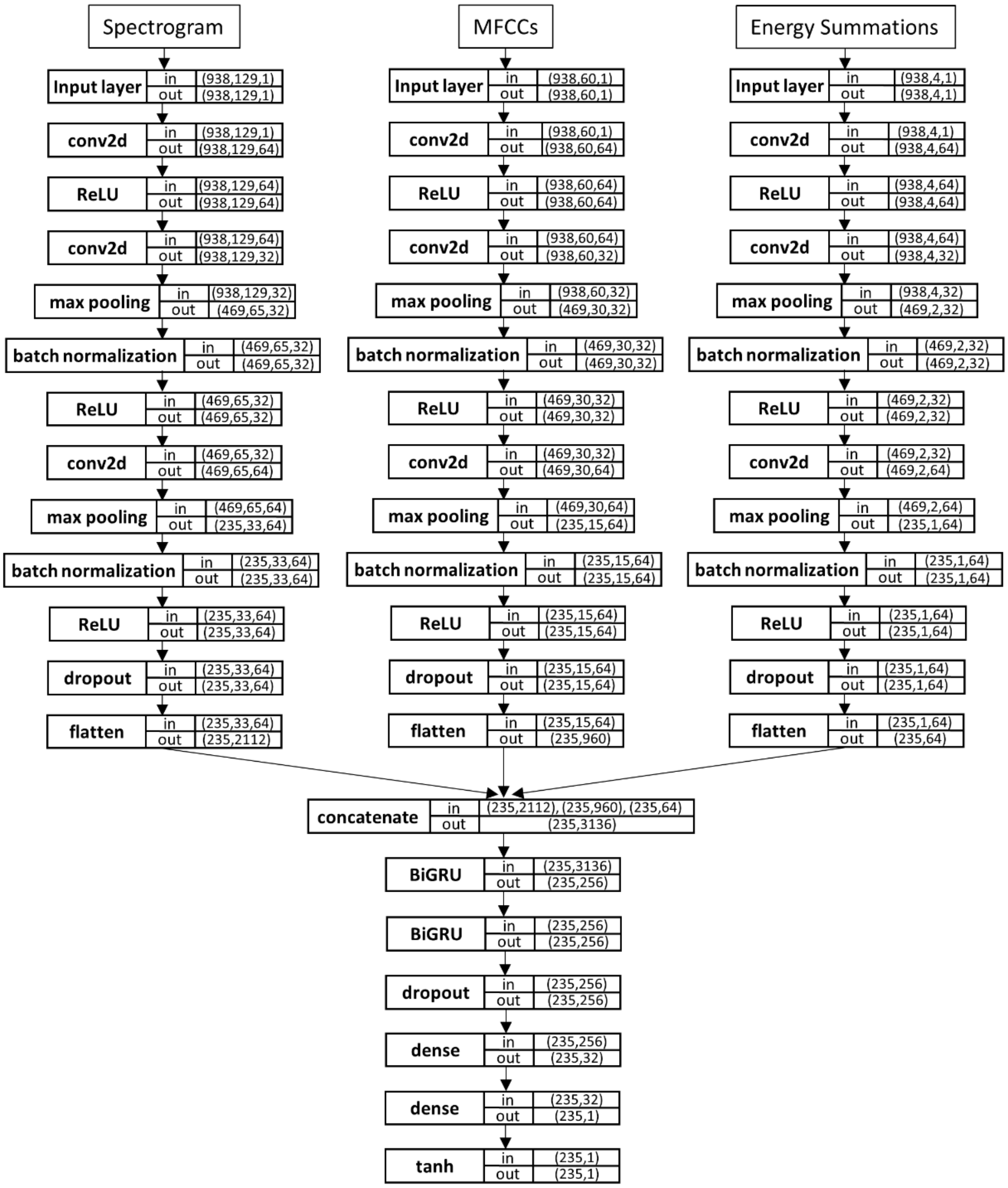

**Figure 5. The architecture of the CNN-BiGRU model.**

The output of the CNN-BiGRU model was an n × 1 vector (n = 235). Each element of the vector represented a time segment in the 15-s. Thresholding was used to classify the output of the tanh layer, $X$: $x \in [-1, 1]$, into three classes, $Y$: $y \in$ {inhalation, background, exhalation} (see eq. 1):

$$Y(x) = \begin{cases} inhalation, if\ x \geq p \\ background, if\ q < x < p \\ exhalation, if\ x \leq q \end{cases} \quad (1)$$

A time segment was classified into either inhalation, exhalation, or background.

4.8 Postprocessing

The segment-prediction results (Fig 6a) were further postprocessed to produce the event-prediction results. First, we merged adjacent time segments into events if they shared the same assigned label (Fig 6b). Then, if any two events were only spanned by a short interval (≤*u* sec) and their maximum energy peaks were within a very close frequency range (Δ25 Hz), we further merged them into a single event (Fig 6c). After the merging, we further removed the burst events (duration <*v* sec) (Fig 6d). The *u* and *v* were set to 0.35 and 0.35, respectively.

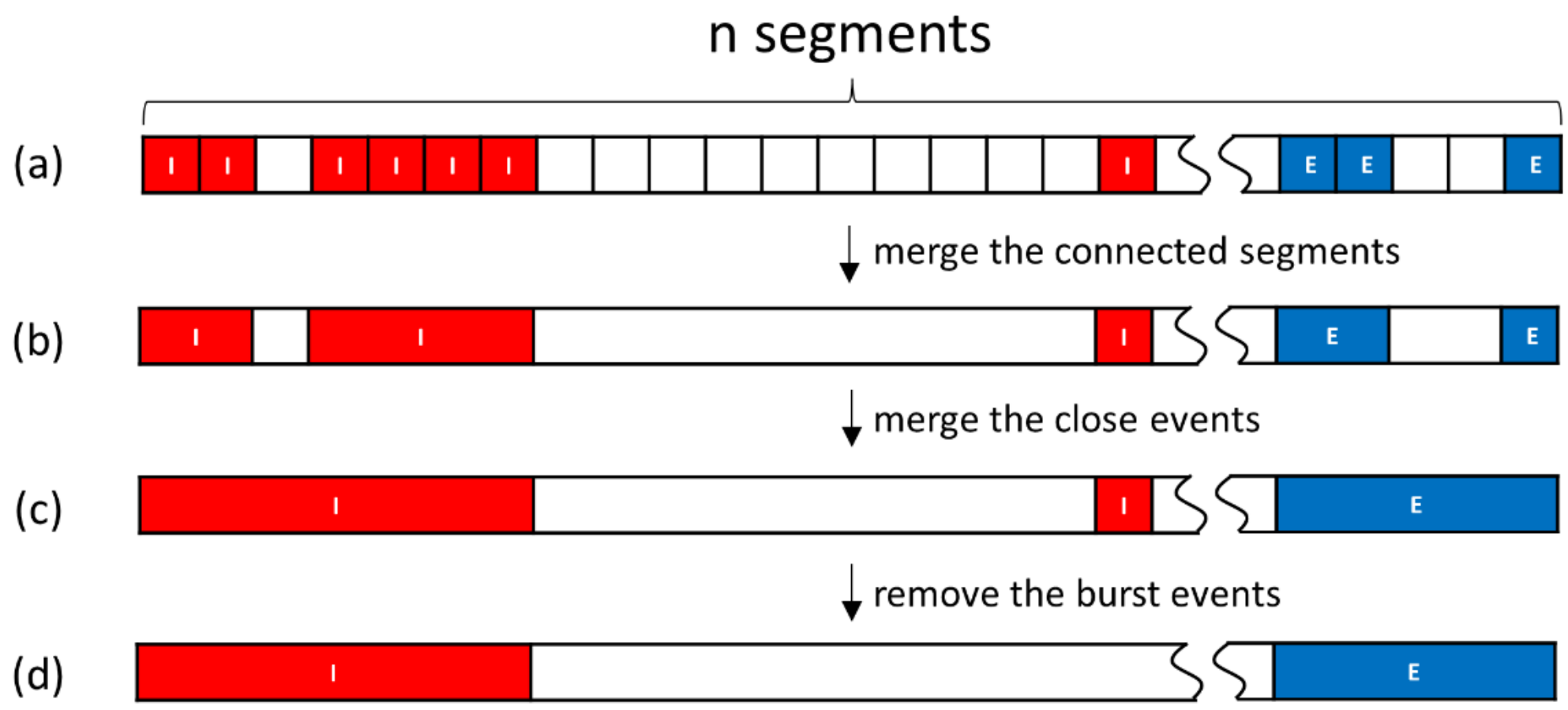

**Figure 6. The flowchart of postprocessing.** (a) the results of segment prediction, (b) the connected segments were merged, (c) the neighboring events (the span was ≤*u* sec) were merged, and (d) the burst events (if duration <*v* sec) were removed. Red color blocks with I stand for inhalations. Blue color blocks with E stand for exhalations.

### 4.9 Training Environment and Parameters

We used a batch size of 64 and an initial learning rate of 0.0001 to train the CNN-BiGRU model using the Adam optimizer. Mean squared error was used as the cost function. The number of epochs was set to 5000, and early stopping was used with a patience of 50. The model was trained on a cloud server (OS: Ubuntu 18.04; CPU: Intel Xeon Gold 6154@3.00 GHz; RAM: 90 GB) provided by Taiwan Web Service. We used TensorFlow 2.1.0 as the framework for the deep neural networks. GPU acceleration was run on an NVIDIA Titan V100 GPU card based on the CUDA 10 and CuDNN 7 frameworks.

## 5. Prediction Refinement and Concatenation

Supposedly, the proposed device executes an inference every *d* seconds. Therefore, the signals from the $(i\text{-}1)d$ to $15+(i\text{-}1)d$ contained in the 15-sec inference windows, $W_{i=1,2...}$ (Fig 7a), were fed into the CNN-BiGRU model in sequence. Fig 7 illustrates how the proposed device generated a prediction result in 15-s windows, refined the prediction results from sequential 15-sec windows, and concatenated the detected inhalation events into an expansive history list. The proposed device did not specifically deal with the detected exhalations because the respiratory rate could be derived as long as inhalations were detected. Hence, the prediction refinement was only applied to detected inhalations. The prediction refinement included the three processes: 1) incomplete event removal, 2) false positive removal, and 3) double count removal, which are introduced in the following sections.

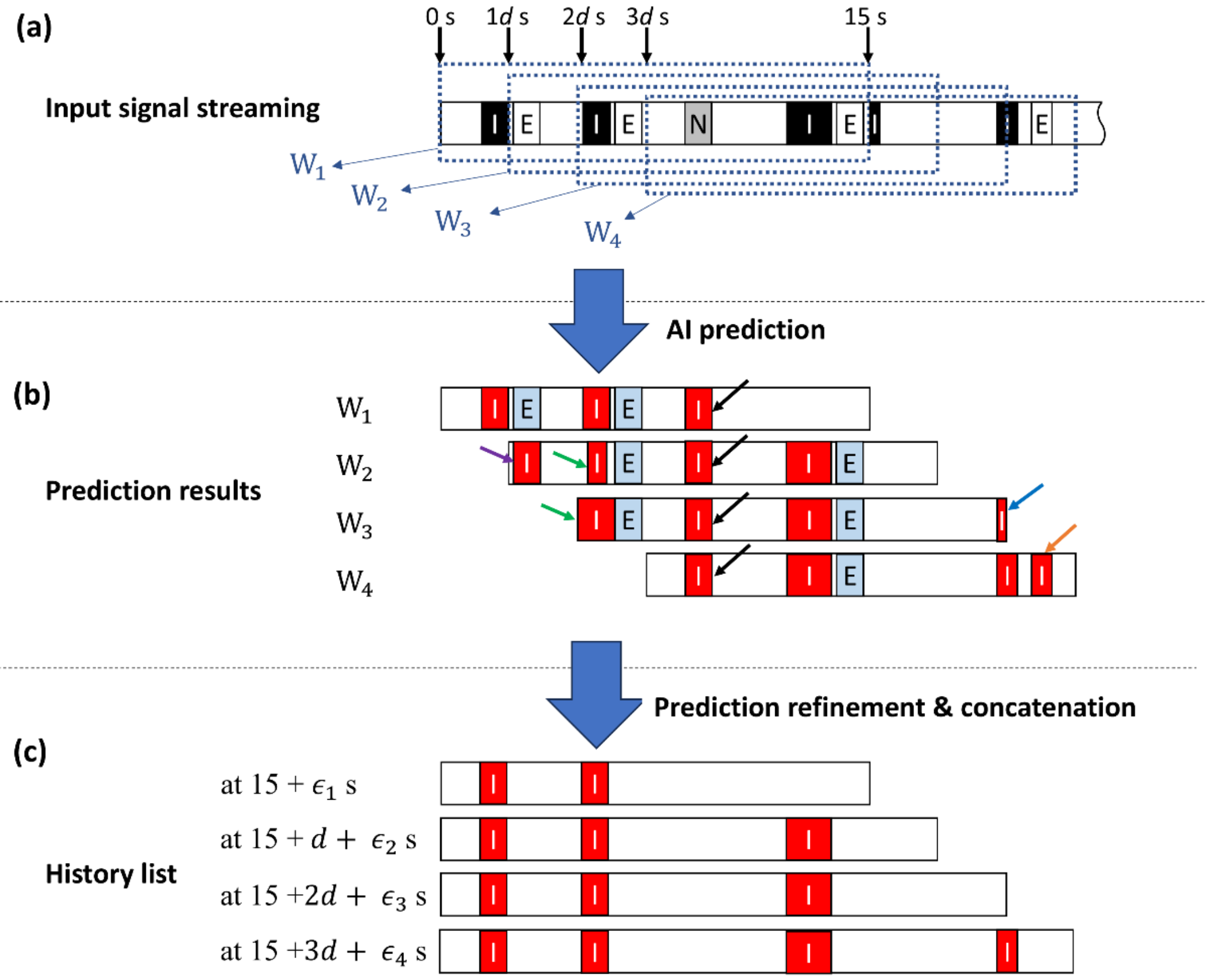

**Figure 7. Steps of prediction refinement and concatenation.** (a) Input signal streaming: black I blocks and white E blocks represent real inhalation and exhalation events; gray N blocks mimic respiratory noise. Sequential 15-second windows (blue dashed squares) are processed by the AI prediction module. (b) Detection results: red I blocks are detected inhalations, sky-blue E blocks are detected exhalations. The purple arrow shows an exhalation misclassified as an inhalation. Green arrows indicate variable inhalation durations. Black arrows mark false positives. Blue arrow shows an incomplete event. The orange arrow shows an exhalation misidentified as an inhalation due to a long inspiratory pause. (c) Refined inhalation events (red blocks) written to the history list at times $15+\epsilon_1$, $15+d+\epsilon_2$, $15+2d+\epsilon_3$, and $15+3d+\epsilon_4$, where $\epsilon_i$ represents processing latency.

## 5.1 Incomplete Event Removal

Sometimes, an ongoing inhalation has not entirely ended, and only a part of it is captured in a 15-sec inference window. Thus, the duration of the detected event (blue arrow in the $W_2$ in Fig 7b) would be much shorter than the actual one. To avoid this situation, we erased the detected inhalation events whose start times are not located in the first 14 seconds of the 15-sec window. After applying the incomplete event removal, the same event is not written into the history list until the $W_4$ (Fig 7c).

5.2 False Positive Removal

To prevent false positives during quiet periods, we calculate the average energy (125–500 Hz) of each predicted inhalation event. Events with energy below a threshold are discarded, as shown in Fig. 7: noise mistakenly identified as inhalations (black arrows in Fig. 7b) is filtered out (Fig. 7c).

5.3 Double Count Removal

The CNN-BiGRU model heavily relied on recognizing a regular inhalation-exhalation-expiratory pause breathing pattern (Fig 8a) to work. If the breathing did not follow such a pattern, e.g., when the inspiratory pause is too long (Fig 8b), the model may wrongly identify the exhalation as an inhalation (orange arrow in the $W_4$ in Fig 7b). We called this condition a double count. The double count doubles the estimated respiratory rate, so it must be eliminated.

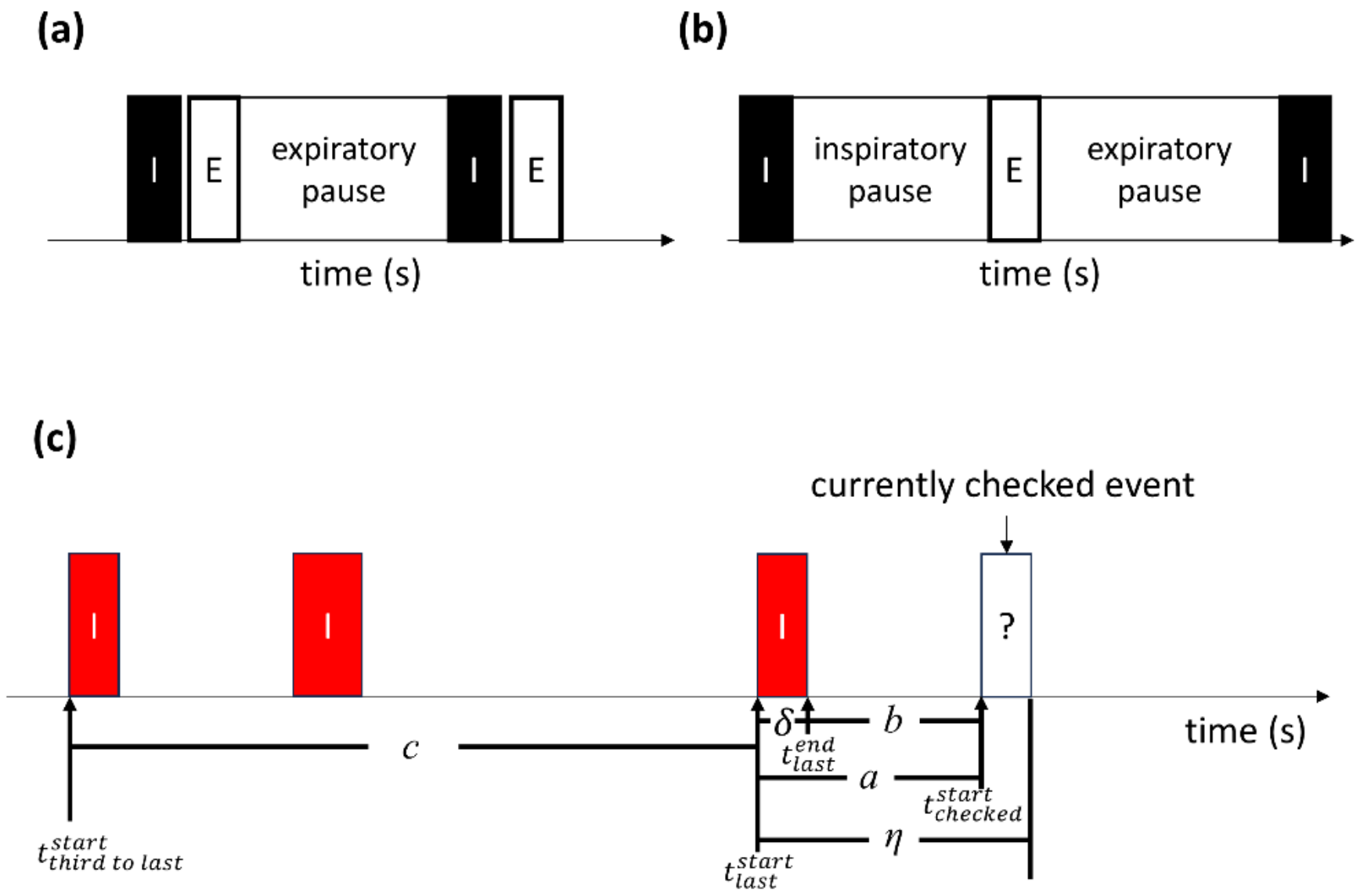

**Figure 8. (a) a regular inhalation-exhalation-expiratory pause breathing pattern, (b) an unusual breathing pattern in which the inspiratory pause is long, and (c) an illustration of parameters in double count removal.** Black I blocks stand for actual inhalations. White E blocks stand for actual exhalations. Red I blocks stand for detected inhalation events already written in the history list. Based on the 1:1 inhalation-to-exhalation ratio in voluntary breathing, we designed a double count removal strategy. We defined four parameters: a (interval between start times of current and last events), b (interval between start time of current event and end time of last event), c (time span between third-to-last and last event start times), and δ (duration of last event). We calculated the average breathing period $\tau = c/2$ and current respiratory rate $RR_current = 60/\tau$. Using the function $\pi(\delta, RR_current)$, we computed the ratio σ to determine the unresponsive period η. A newly predicted inhalation event is removed if any of these conditions are met: (1) $a < 1$, (2) $b < 0.3$, (3) $a < 3$ and $a < \tau/2$, or (4) $t_checked_start < (t_last_start + \eta)$. These parameters were determined empirically.

5.4 Concatenation of Prediction Refined Results

When the proposed device starts running, an empty history list is initiated. As the model identified the inhalation events in each sequential 15-sec window and refined, the start and end times of the inhalation events were concatenated to the history list. To not allow the same inhalation with minor variations (the green arrows in the $W_2$ and $W_3$ in Fig 7b) in start and end times to be repeatedly written into the history list and not allow an inhalation (the purple arrow in the $W_2$ in Fig 7b) to be inserted before an event already existed in the history list that confused respiratory rate estimation.

6. Respiratory Rate Estimation

The proposed device estimates respiratory rate on the basis of the average period of the latest three breaths, as shown in the equation below:

$$Respiratory\ Rate = \frac{60}{mean\ of\ periods\ of\ the\ last\ 3\ breaths}.$$

The period of breath was defined as the difference between the start times of two consecutive inhalation events. The estimation process is illustrated in Fig 9. The proposed device does the 1st, 2nd, 3rd, …, and *n*th estimation as time continues to proceed. In the 1st estimation cycle, we have not had three inhalations (2 breath periods) written in the history list, so the respiratory rate estimation does not start.

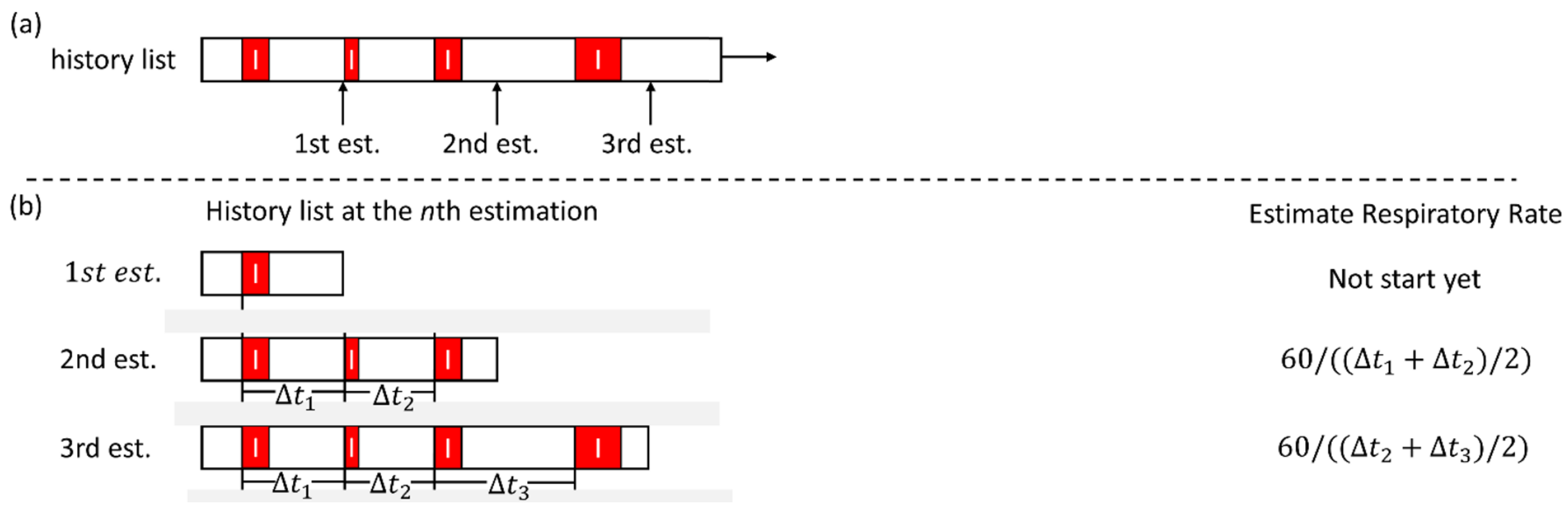

**Figure 9. Illustration of an example for respiratory rate estimation.** (a) The expanding history list, (b) the history lists observed at the 1st, 2nd, …, and 6th estimations are displayed on the left; the respiratory rate is shown on the right. The red blocks represent inhalation events registered in the history list.

7. AI Performance Evaluation

In this study, we reported the model performance of inhalation and exhalation event detection. Please refer to our previous study for the definition of true positive, false positive, and false negative events (23). We tested three prediction intervals (2, 5, and 15 s) to simulate the proposed device facing continuous tracheal sound streaming. We defined two conditions for testing the effect of prediction refinement for inhalation detection: 1) the prediction results of sequential 15-s files were concatenated based on the concatenation rules, but the three prediction refinement processes were not activated, and 2) the prediction results went through all the prediction refinement processes first, and then the refined results were concatenated. As for exhalation detection, the detected events were concatenated based on the same rule, but prediction refinement was not applied. In real-world use, the proposed device did not record the detected exhalations.

The sensitivity (SEN), positive predictive value (PPV), and F1-score of event detection were reported when the thresholds were set as 0.5 and -0.5 for inhalation and exhalation, respectively. The respiratory rate estimation was validated against capnometry in a multi-center clinical trial, which will be published in another paper.

## 8. Results

**Table 2. Characteristics of tracheal sound recordings and labels in the HF_Tracheal_V2.**

| Database | HF_Tracheal_V2 | | |
|---|---|---|---|
| Data set | Training | Test | Total |
| n of recordings | 311 | 62 | 373 |
| total duration (min) | 6689.9 | 1831.4 | 8521.3 |
| Inhalation | | | |
| n | 82275 | 25369 | 107644 |
| total duration (min) | 1574.3 | 513.5 | 2087.8 |
| mean duration (sec) | 1.15±0.33 | 1.21±0.37 | 1.16±0.35 |
| Exhalation | | | |
| n | 77778 | 24868 | 102646 |
| total duration (min) | 1471.2 | 465.2 | 1936.4 |
| mean duration (sec) | 1.13±0.53 | 1.12±0.50 | 1.13±0.52 |

Table 2 lists the characteristics of tracheal sound recordings and labels in HF_Tracheal_V2. After truncating the files in the training set using a 15-s window with a 5-s stride for data augmentation, the number of the I increased from 82275 to 262905 and the number of E labels increased from 77778 to 251694, respectively.

**Table 3. Performance of the CNN-BiGRU model.**

| Task | Prediction Refinement | Prediction Interval (sec) | Event | | |
|---|---|---|---|---|---|
| | | | SEN | PPV | F1 Score |
| Inhalation | Off | 2 | 0.861 | 0.823 | 0.842 |
| | Off | 5 | 0.858 | 0.841 | 0.850 |
| | Off | 15 | 0.754 | 0.829 | 0.790 |
| | On | 2 | 0.828 | 0.832 | 0.830 |
| | On | 5 | 0.828 | 0.848 | 0.838 |
| | On | 15 | 0.728 | 0.831 | 0.776 |
| Exhalation | NA | 2 | 0.836 | 0.822 | 0.829 |
| | NA | 5 | 0.833 | 0.841 | 0.837 |
| | NA | 15 | 0.727 | 0.834 | 0.777 |

SEN: sensitivity, and PPV: positive predictive value.

Table 3 tabulates the performance of the CNN-BiGRU model in detecting inhalations and exhalations in the test data. The results show that turning on the prediction refinement in inhalation event detection can get a better PPV and a poorer SEN, which yet consequently results in a poorer F1 score. For the prediction interval, setting it to 5 s improves PPV and F1 for inhalation and exhalation detection.

Fig 10a-c displays the 30-s spectrograms of the main, env, and audio signals of a tracheal sound recorded by the proposed device from a subject lying on a bed in a noisy environment. We can observe the env signal (Fig 10b) contained background noise (blue arrows) plus some human voice (white arrows) that blurred the breathing sound patterns in the main signal (Fig 10a). After the ANC process, the proposed device can broadcast less-interfered auscultated sound and display clearer breathing sound patterns in the spectrogram (Fig 10c).

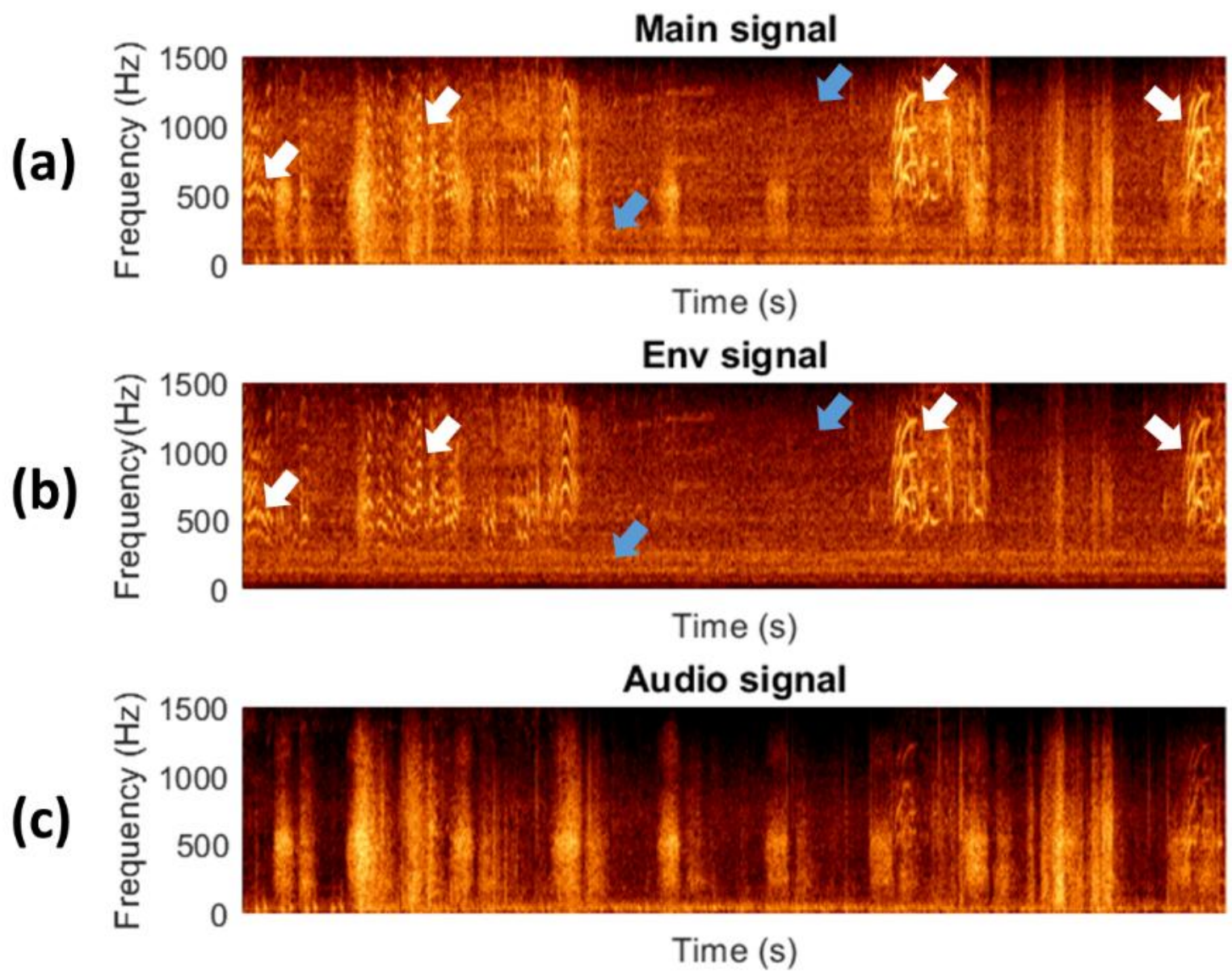

**Figure 10. The spectrograms of the (a) main, (b) env, and (c) audio signals of a tracheal sound recorded by the proposed device from a subject in a noisy environment.** White arrows indicate background human voice, and blue arrows indicate other background noise.

Fig 11 shows the proposed device used to monitor a subject receiving sedation during a procedure. The anesthesiologist can immediately notice the subtle change in the subject's airway condition by hearing the tracheal sound broadcast by the proposed device or by observing the real-time display of the spectrogram.

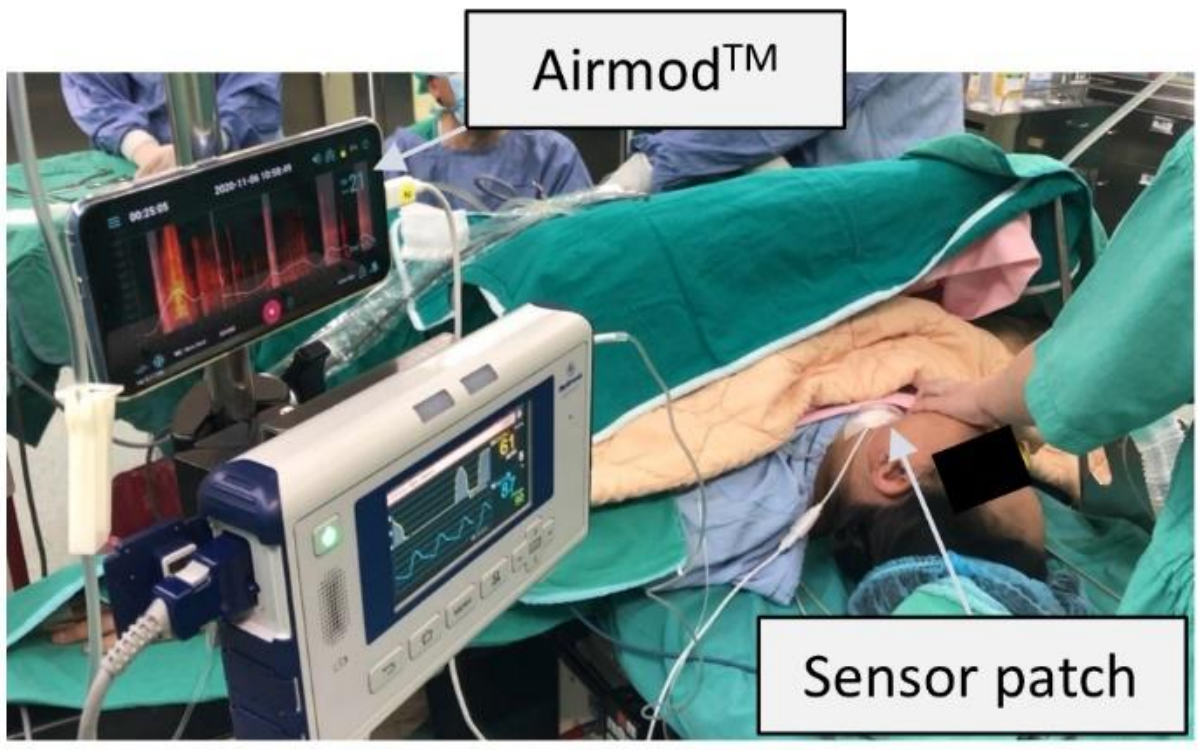

**Figure 11. The proposed device was used for tracheal sound monitoring in the clinic.** The acoustic sensor of the electronic stethoscope was attached to the pretracheal region of a subject.

Fig 12 illustrates the spectrogram, detected inhalations, and the curve of the estimated respiratory rate of a 90-s tracheal sound signal recorded by the proposed device and re-drawn for publication purposes. We observe the spectrograms (top graphs in Fig. 12a-c) and see that the subject breathed stably at the beginning and increased breathing speed during 18–30 sec; the subject stopped breathing during 30–64 sec and resumed breathing after 64 sec. The proposed device successfully detected all the inhalations (middle graphs in Fig12a and c) and estimated the breath-to-breath respiratory rate (black curve in the bottom graphs in Fig12a and c). When no inhalations were detected for a while and the estimated respiratory rate fell below 4 bpm (blue arrow in Fig. 12b and c), the proposed device did not display a value.

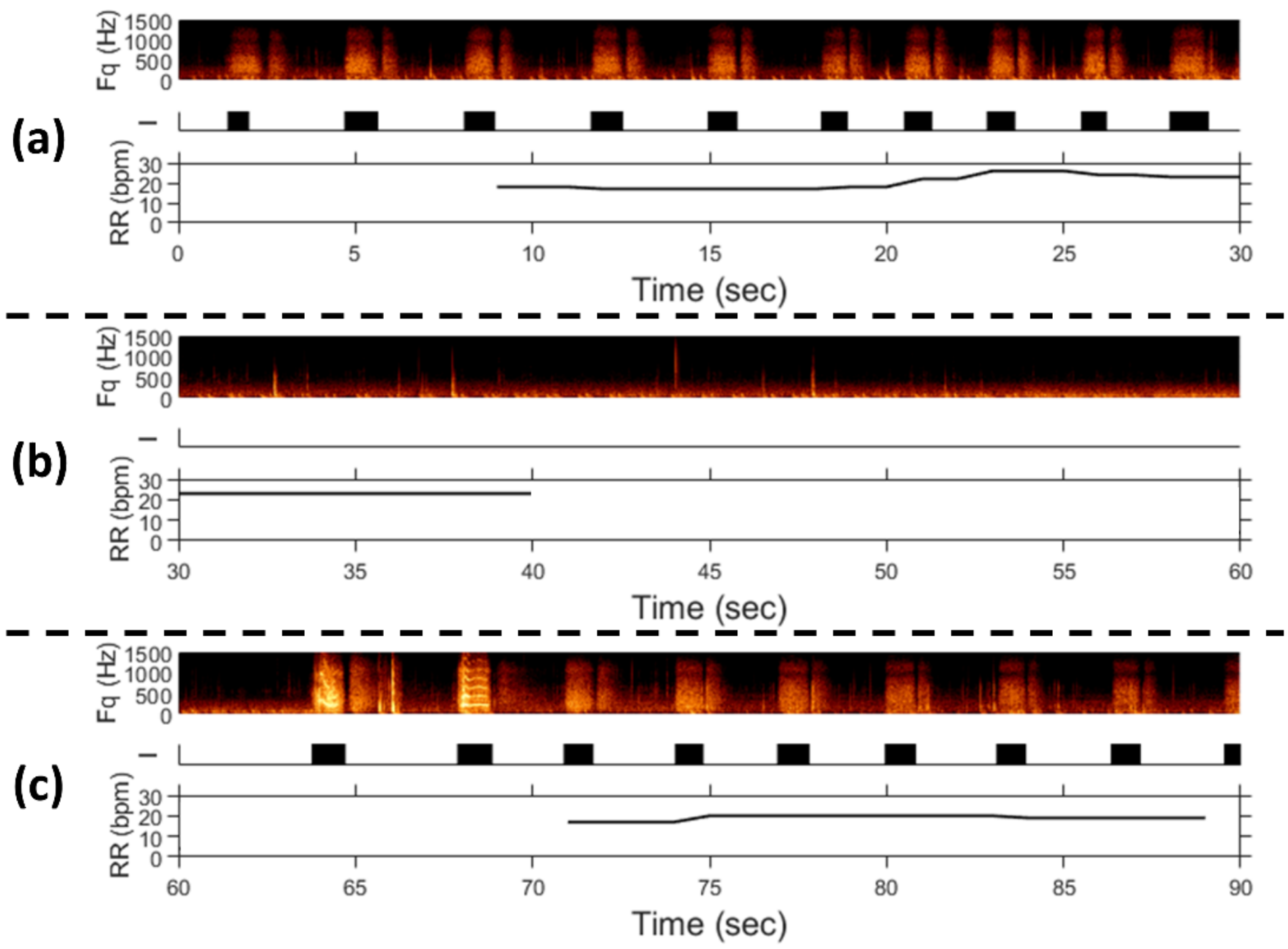

**Figure 12. The spectrogram, detected inhalations, and the curve of estimated respiratory rate in (a) 0–30 s, (b) 30–60 s, and (c) 60–90 s in a tracheal sound signal.** In (a)–(c), the top graph

displays the spectrogram, the middle graph shows the detected inhalations (black blocks), and the bottom graph displays the estimated respiratory rate (black curve). Fq: frequency, I: inhalation, and RR: respiratory rate.

9. Discussion

The accuracy of respiratory rate estimation in the proposed device hinges on the performance of the CNN-BiGRU model. Table 3 shows the model performs adequately on inhalation and exhalation detection in the tracheal sounds. We can observe that setting the prediction interval shorter (from 15 sec to 5 sec or from 15 sec to 2 sec) can improve performance. This resulted from events being truncated in half at the very beginning and end of each sequential 15-s window, which could be viewed as incorrect detection (JI < 0.5). The wrong detection is likely to be reduced when the truncated events are shifted to the middle of the subsequent inference windows. Table 3 also shows that the model had a lower PPV and F1-score, and a slightly higher SEN, at a prediction interval of 2 sec compared with 5 sec. It is because if a ground-truth event is not detected in the previous predictions, the model can have more opportunities to detect it in the subsequent inference windows when the prediction interval is set shorter; however, from another angle, a false positive event is more likely to pop up in so many times of predictions. Therefore, it is a trade-off between a higher F1 score and a shorter prediction interval. Because clinicians cannot tolerate a 5-sec delay in prediction, the default prediction interval is set to 2 sec in the proposed device. Furthermore, the results indicate that prediction refinement traded off the SEN of inhalation detection against PPV, thereby lowering the overall F1-score. Because a false positive event poses a higher risk than a false negative event, we have the prediction refinement activated by default to pursue a higher PPV.

Fig. 10 shows that the ANC in the proposed device significantly improves the distinguishability of inhalation and exhalation. Theoretically, the ANC technology can attenuate the broadband ambient noise to a certain level (-19 dB at 100−400 Hz in personal active noise reduction devices (34)) when the filter order is fixed and limited. Thus, if the ambient noise is not too loud or its frequency range falls within the ANC's operating range, the auscultated sound and the displayed spectrogram would not be affected. However, the ANC can’t work on the sound picked up only by the main mic, not by the env mic. Therefore, instrumental noise caused by touching the acoustic sensor or by rubbing the acoustic cable against a surface, and noise generated in the subject's airway, such as the hissing of supplemental oxygen flow, coughing, and vocalization, can still influence the performance of the proposed device.

Currently, the proposed device must run on a smartphone (ROG5, Asus, Taipei, Taiwan) or on other devices with powerful CPUs. This is because the CNN-BiGRU model drained the Android device's computational resources. GPU acceleration is not used because the power charging speed can’t match the power consumption rate with a high-speed charging feature. The model makes it impossible to port the proposed device to a cheaper medical-grade personal computer. Our other preliminary studies (not published) suggest that the You Only Look Once (YOLO) object detection model v5 (35) can perform respiratory sound analysis with high efficiency and achieve comparable, if not better, performance. In the future, we will validate YOLOv5 and its later variants, or other state-of-the-art models, to replace the CNN-BiGRU model in the proposed device and reduce computational cost.

As of writing this paper, we have completed a multicenter clinical trial to validate respiratory rate estimation against capnometry. The validation results will be published in a separate paper.

## 10. Conclusion

In this paper, we present an AI-powered respiratory sound monitor and describe its key functions and processing steps for estimating breath-to-breath respiratory rate. We demonstrate that the proposed device enables continuous respiratory sound monitoring. The proposed device can replace conventional tracheal sound auscultation, free clinicians' hands, and prevent damage to their ears from sudden loud noise. The ANC technology allows the proposed device to resist ambient noise interference during auscultation. The real-time spectrogram display can further aid the clinician in identifying respiratory abnormalities.

**Acknowledgment**

Wan-Ling Tsai, Nian-Jhen Lin, Jia-Yin Lin, Hui-Chen Teng, and Yu-Chun Yo labeled the tracheal sounds. The quality of labels is reviewed by Ching-Shiang Lu and Chuan Chen. We thank Chun-Chieh Chen and Fang-Shi Wang with Avalanche Computing Inc., who supported the training of the AI model.

**Conflict of Interests**

Shang Ran Huang, Fu-Shun Hsu, Ji-Chang Huang, Chuan Chen, and Yuan-Ren Cheng are/were employees of Heroic Faith Medical Science Co., Ltd. Shou-Zen Fan, Chang-Fu Su, and Eng-Che Huang own stocks of Heroic Faith Medical Science Co., Ltd.

**References**

1. Das S, Ghosh S. Monitored anesthesia care: An overview. Journal of anaesthesiology, clinical pharmacology. 2015;31(1):27.
2. Hillier SC, Mazurek MS. Monitored anesthesia care. Clinical anesthesia 3rd ed Philadelphia: Lippincott-Raven. 1996:1159-71.
3. Kamat PP, McCracken CE, Simon HK, Stormorken A, Mallory M, Chumpitazi CE, et al. Trends in outpatient procedural sedation: 2007–2018. Pediatrics. 2020;145(5).
4. Benzoni T, Cascella M. Procedural Sedation. Treasure Island (FL): StatPearls Publishing; 2022 2022.
5. Nagrebetsky A, Gabriel RA, Dutton RP, Urman RD. Growth of nonoperating room anesthesia

care in the United States: a contemporary trends analysis. Anesthesia & Analgesia. 2017;124(4):1261-7.
6. Yan JW, McLeod SL, Iansavitchene A. Ketamine-propofol versus propofol alone for procedural sedation in the emergency department: a systematic review and meta-analysis. Academic Emergency Medicine. 2015;22(9):1003-13.
7. Newstead B, Bradburn S, Appelboam A, Reuben A, Harris A, Hudson A, et al. Propofol for adult procedural sedation in a UK emergency department: safety profile in 1008 cases. British journal of anaesthesia. 2013;111(4):651-5.
8. Woodward ZG, Urman RD, Domino KB. Safety of non–operating room anesthesia: a closed claims update. Anesthesiology Clinics. 2017;35(4):569-81.
9. Krmpotic K, Rieder MJ, Rosen D. Recommendations for procedural sedation in infants, children, and adolescents. Paediatrics & Child Health. 2021;26(2):128-.
10. Anesthesiologists ASo. Practice guidelines for moderate procedural sedation and analgesia 2018: a report by the American Society of Anesthesiologists Task Force on Moderate Procedural Sedation and Analgesia, the American Association of Oral and Maxillofacial Surgeons, American College of Radiology, American Dental Association, American Society of Dentist Anesthesiologists, and Society of Interventional Radiology. Anesthesiology. 2018;128(3):437-79.
11. Association AD. Guidelines for the use of sedation and general anesthesia by dentists. Adopted by the ADA House of Delegates. 2016.
12. Coté CJ, Wilson S, Pediatrics AAo, Dentistry AAoP. Guidelines for monitoring and management of pediatric patients before, during, and after sedation for diagnostic and therapeutic procedures. Pediatrics. 2019;143(6).
13. Gelb AW, Morriss WW, Johnson W, Merry AF. World Health Organization-World Federation of Societies of Anaesthesiologists (WHO-WFSA) international standards for a safe practice of anesthesia. Anesthesia & Analgesia. 2018;126(6):2047-55.
14. Alabama Administrative Code, 540-X-10-.06 (2003).
15. Arizona Administrative Code, Title 4, Chapter 16, Section 704 (2008).
16. Texas Administrative Code, Title 22, Part 5, Chapter 110, § 5 (2014).
17. California Business and Profession Code, Division 2, Chapter 4, Article 4, § 1682 (2025).
18. McLane I, Emmanouilidou D, West JE, Elhilali M. Design and comparative performance of a robust lung auscultation system for noisy clinical settings. IEEE journal of biomedical and health informatics. 2021;25(7):2583-94.
19. Machado A, Oliveira A, Aparicio J, Marques A. Respiratory auscultation:(dis) agreement between health professionals. Eur Respiratory Soc; 2015.
20. Kim Y, Hyon Y, Lee S, Woo S-D, Ha T, Chung C. The coming era of a new auscultation system for analyzing respiratory sounds. BMC Pulmonary Medicine. 2022;22(1):119.
21. Pramono RXA, Bowyer S, Rodriguez-Villegas E. Automatic adventitious respiratory sound analysis: A systematic review. PloS one. 2017;12(5):e0177926.

22. Kim Y, Hyon Y, Jung SS, Lee S, Yoo G, Chung C, et al. Respiratory sound classification for crackles, wheezes, and rhonchi in the clinical field using deep learning. Scientific reports. 2021;11(1):1-11.
23. Hsu F-S, Huang S-R, Su C-F, Huang C-W, Cheng Y-R, Chen C-C, et al. A dual-purpose deep learning model for auscultated lung and tracheal sound analysis based on mixed set training. Biomedical Signal Processing and Control. 2023;86:105222.
24. Huq S, Moussavi Z. Acoustic breath-phase detection using tracheal breath sounds. Medical & biological engineering & computing. 2012;50:297-308.
25. Kuo SM, Morgan DR. Active noise control: a tutorial review. Proceedings of the IEEE. 1999;87(6):943-73.
26. Turk E, Ulusar UD, Ogunc G, Canpolat M, Yaprak M, editors. Active Noise Cancellation for IoT-Driven Electronic Stethoscope: A Comparative Study of Adaptive Filters. Forthcoming Networks and Sustainability in the IoT Era: First EAI International Conference, FoNeS–IoT 2020, Virtual Event, October 1-2, 2020, Proceedings 1; 2021: Springer.
27. Aviles-Solis JC, Storvoll I, Vanbelle S, Melbye H. The use of spectrograms improves the classification of wheezes and crackles in an educational setting. Scientific Reports. 2020;10(1):1-8.
28. Patel SB, Callahan TF, Callahan MG, Jones JT, Graber GP, Foster KS, et al. An adaptive noise reduction stethoscope for auscultation in high noise environments. The Journal of the Acoustical Society of America. 1998;103(5):2483-91.
29. Hsu F-S, Huang C-J, Kuo C-Y, Huang S-R, Cheng Y-R, Wang J-H, et al., editors. Development of a respiratory sound labeling software for training a deep learning-based respiratory sound analysis model. International Forum on Medical Imaging in Asia 2021; 2021: SPIE.
30. Pasterkamp H, Brand PL, Everard M, Garcia-Marcos L, Melbye H, Priftis KN. Towards the standardisation of lung sound nomenclature. European Respiratory Journal. 2016;47(3):724-32.
31. Boriosi JP, Zhao Q, Preston A, Hollman GA. The utility of the pretracheal stethoscope in detecting ventilatory abnormalities during propofol sedation in children. Pediatric Anesthesia. 2019;29(6):604-10.
32. Berry RB, Abreu AR, Krishnan V, Quan SF, Strollo Jr PJ, Malhotra RK. A transition to the American Academy of Sleep Medicine–recommended hypopnea definition in adults: initiatives of the Hypopnea Scoring Rule Task Force. Journal of Clinical Sleep Medicine. 2022;18(5):1419-25.
33. Ranjan R, Thakur A. Analysis of feature extraction techniques for speech recognition system. International Journal of Innovative Technology and Exploring Engineering. 2019;8(7C2):197-200.
34. Rudzyn B, Fisher M. Performance of personal active noise reduction devices. Applied Acoustics. 2012;73(11):1159-67.
35. Jocher G, Stoken A, Chaurasia A, Borovec J, Kwon Y, Michael K, et al. ultralytics/yolov5: v6. 0-YOLOv5n'Nano'models, Roboflow integration, TensorFlow export, OpenCV DNN support. Zenodo. 2021.